\documentstyle[aps,prl,epsf]{revtex}

\newcommand{\etal}{$\it et~al.$}

\newcommand {\Fig}[1]{Fig.$\;$\ref{#1}}
\newcommand {\Eq}[1]{Eq.$\;$\ref{#1}}

\newcommand {\be} {\begin{equation}}
\newcommand {\ee} {\end{equation}}

\newcommand{\br} {{\bf r}}

\begin{document}
%\DAvance\textheight by 0.5in
%\DAvance\topmargin by -0.2in
%\setlength{\baselineskip}{24pt}

%\twocolumn[\hsize\textwidth\columnwidth\hsize\csname@twocolumnfalse%
%\endcsname

\title{Phase Field Methods and Dislocations}

\author{D. Rodney and A. Finel}

\address{Laboratoire d'Etude des Microstructures, CNRS-ONERA, B.P. 72, 92322 Chatillon Cedex, France}

\date{\today}
\maketitle

%\tableofcontents

\begin{abstract}

We present a general formalism for incorporating dislocations in Phase Field methods. This formalism is based on the elastic equivalence between a dislocation loop and a platelet inclusion of specific stress-free strain related to the loop Burgers vector and normal. Dislocations are thus treated as platelet inclusions and may be coupled dynamically to any other field such as a concentration field. The method is illustrated through the simulation of a Frank-Read source and of the shrinkage of a loop in presence of a concentration field.

\end{abstract}

\vspace{.5truein}
%]

%\twocolumn
%\narrowtext

\section{INTRODUCTION}

Phase Field methods (PFM) are powerful tools to study microstructural evolutions during phase transformations at a mesoscopic space scale of typically 1 $\mu$m \cite{tms}. These methods account for both chemical and elastic contributions to the total energy. The chemical contribution is usually modeled by means of a coarse-grained Ginzburg-Landau free energy while elastic coherency strains are accounted for by recourse to Khachaturyan's Phase Field Microelasticity theory \cite{khachaBook}.

Up to now, the question of the influence of crystalline defects, such as dislocations or grain boundaries, on phase transformations has rarely been addressed in the context of PFMs, although such defects are known to strongly influence phase transformations, since, for example, they are preferential sites for the nucleation of precipitates \cite{larche}. The first attempt to introduce dislocations in a PFM is due to L\'eonard and Desai \cite{leonard} who studied the influence of dislocations on the spinodal decomposition of a binary alloy. Their model is limited to straight edge or screw dislocations in a 2D elastically isotropic medium. Moreover, the dislocations are static. Spinodal decomposition in presence of dislocations has also been considered in a PFM by Hu and Chen \cite{chen}.

On the other hand, the dynamics of 3D dislocation microstructures may be simulated at the same mesoscopic scale as that of the PFMs using Discrete Dislocation Dynamics (DDD) codes \cite{ddd}. Up to now, the presence of misfitting inclusions has been accounted for in these methods only by means of a background coherency stress, not evolving with time .

We present in this article a general formalism for incorporating dislocations in Phase Field Methods with a dynamical coupling between dislocations and a phase transformation \cite{finel}. This method is valid for any 3D anisotropic elastic medium with any dislocation distribution. A similar method is being presently developed by Wang and co-workers \cite{KhachaActaMet}. It is based on an analogy, first pointed out by Nabarro \cite{nabarro}, between a planar dislocation loop and a platelet inclusion with a stress-free strain related to the Burgers vector {\bf b} and normal {\bf n} of the loop by $\epsilon^{oo}_{ij} = (b_i n_j + b_j n_i)/2d$, where $d$ is the interplanar distance. The basic idea of the method is therefore to introduce dislocations as a new phase made of platelet inclusions. Dislocations may thus be treated in a PFM using the same formalism as the one developed for phase transformations. Moreover, as will be seen below, kinetic equations adapted to dislocation motion can be introduced, leading to a coupling between the dynamics of the dislocations and that of a phase transformation.

\section{STATIC MODEL}

We present in this section the method to incorporate static dislocations in a Phase Field Method.

\subsection{The Phase Field Microelasticity theory}

The model is based on the Phase Field Microelasticity theory which is briefly summarized here.

We consider a homogeneous elastic medium with elastic tensor $\lambda_{ijkl}$, subjected to an external uniform stress $\sigma^A_{ij}$. Let $\epsilon^0_{ij}({ \bf r})$ be the stress-free strain at point \br. The microstructure is composed of several phases, noted $p$, characterized by their density field $\theta_p({\bf r})$ and their stress-free strain $\epsilon^{00}_{ij}(p)$. The total stress-free strain is the sum of the stress-free strains of the various phases:

\be
\epsilon^0_{ij}({\bf r}) = \sum_{p:~phases} \epsilon^{00}_{ij}(p) \theta_p({\bf r})
\ee

The elastic energy in a system of volume V is:

\begin{eqnarray}
E^{el} = \frac{1}{2} \int_V d^3r~\lambda_{ijkl} (\epsilon_{ij}(\br) - \epsilon^0_{ij}(\br) ) (\epsilon_{kl}(\br) - \epsilon^0_{kl}(\br) )
- \int_V d^3r~\sigma^A_{ij} \epsilon_{ij}(\br)
\label{Eelastic}
\end{eqnarray}

Following Khachaturyan \cite{khachaBook}, the equilibrium state of the system is obtained analytically in reciprocal space, by decomposing the strain field into homogeneous and inhomogeneous strain fields and imposing periodic boundary conditions to the inhomogeneous strain field. The equilibrium elastic energy may then be written as a functional of the Fourier transforms of the density fields:

\begin{eqnarray}
E^{el} = \frac{V}{2} \sum_{{\bf K} \neq 0} \sum_{p,q:~phases} B_{pq}({\bf K}) \theta_p({\bf K}) \theta_q^{\ast}({\bf K})
- V \sum_{p:~phase} \sigma^A_{ij} \epsilon^{00}_{ij}(p) \theta_p({\bf K=0})
\label{EFinale}
\end{eqnarray}
with 
\be
B_{pq}({\bf K}) = \sigma^{00}_{ij}(p) \epsilon^{00}_{ij}(q) - K_i \sigma^{00}_{ij}(p) G_{jm}({\bf K}) \sigma^{00}_{mn}(q) K_n
\ee

The tensor $G_{im}({\bf K})$ is the Fourier transform of Green's function (defined by $G^{-1}_{im}({\bf K}) = \lambda_{ijml} K_j K_l$) and $\sigma^{00}_{mn}(p)$ is related to the stress-free strain by Hooke's law. The matrices $B_{pq}({\bf K})$ are intrinsic quantities which characterize the elastic interactions between phases $p$ and $q$.

We mention for later use that the equilibrium displacement field is given by:

\be
u_i({ \bf r}) =  \overline{\epsilon}_{ij} r_j + \sum_{{\bf K} \neq 0} -i G_{im}({\bf K}) \sigma^0_{mn}({\bf K}) K_n \exp{i {\bf K}{\bf r}}
\label{deltaueq}
\ee
where $\overline{\epsilon}_{ij}$ is equal to the spatial average of the stress-free strain and $\sigma^0_{mn}({\bf K})$ is the Fourier transform of the stress related to the stress-free strain by Hooke's law.

\subsection{Application to dislocations}

\subsubsection{Equivalence between a dislocation loop and a platelet inclusion}

The method is based on the elastic equivalence between a dislocation loop of Burgers vector {\bf b} and normal {\bf n} and a platelet inclusion, bounded by the dislocation line, of vanishing thickness $d$ and of stress-free strain:

\be
\epsilon^{oo}_{ij} = \frac{1}{2} \frac{b_i n_j + b_j n_i}{d}
\label{stressfreestrain}
\ee
and density field $\theta({\bf r}) = 1$ in the inclusion and $\theta({\bf r}) = 0$ outside.

Inclusion and dislocation loop are elastically equivalent because they produce the same equilibrium displacement field, given by the well-known Burgers formula \cite{landau}. To show this, we note that, in the limit of vanishing thickness $d$, the Fourier transform of the inclusion density field becomes a surface integral:

\begin{eqnarray}
\theta({\bf K})  = \frac{1}{V} \int_V d^3 r~\theta({\bf K}) \exp{-i {\bf K}{\bf r}}
 \rightarrow \frac{d}{V} \int_S dS \exp{-i {\bf K}{\bf r}}
\label{thetak}
\end{eqnarray}

From \Eq{deltaueq}, using the definition of $\sigma^0_{mn}({\bf K})$ and the symmetries of the elastic tensor, the equilibrium displacement field produced by a single platelet inclusion in the limit of infinite volume ($\overline{\epsilon}_{ij} = 0$) is:

\be
u_i({\bf r}) = -i \lambda_{mnkl} \frac{b_k n_l}{d} \sum_{{\bf K} \neq 0} G_{im}({\bf K}) \theta({\bf K}) K_n \exp{i {\bf K}{\bf r}}
\ee

Using \Eq{thetak} in the limit of vanishing thickness, we obtain  Burgers formula:

\begin{eqnarray}
u_i({\bf r}) &=& -i \lambda_{mnkl} \frac{b_k n_l}{V} \int_S dS' \sum_{{\bf K} \neq 0} G_{im}({\bf K}) K_n \exp{i {\bf K}({\bf r}-{\bf r'})} \nonumber \\
           &=& \lambda_{mnkl} b_k n_l \int_S dS' \frac{\partial}{\partial r'_n} G_{im}({\bf r}-{\bf r'}) 
\end{eqnarray}
where $G_{im}(\br) = \int \frac{d^3 k}{(2 pi)^3} G_{im}({\bf K}) \exp{ i {\bf K} {\bf r} }$ is the real space representation of Green's function in an infinite volume.

This equivalence allows us to treat simply dislocations in a Phase Field method as inclusions of an extra phase. The method is general in the sense that the medium may be elastically isotropic or anisotropic and that no constraint is imposed on the character of the dislocations, which can be edge, screw or mixed and can be shear loops as well as vacancy or interstitial loops.

\subsubsection{Case of crystalline solids}

In crystalline solids, dislocations belong to a finite set of slip systems, i.e. they may adopt only a finite number of Burgers vectors and slip planes dictated by the crystallography of the material. For example, there are 12 slip systems in FCC systems, made of $a/2 \langle 110 \rangle$ Burgers vectors and $\{ 111 \}$ slip planes.

In the context of the present model, configurations with several slip systems can be modeled by associating one field to each slip system. A given slip system is then characterized by a stress-free strain given by \Eq{stressfreestrain} and a density field $\theta({\bf r})$ equal to the total slip at point \br~produced by the motion of the dislocations of the slip system. The density fields may therefore take any positive or negative integer value depending on the sign and number of dislocations which have glided over point \br.

Multislip deformation is therefore modeled by means of a total stress-free strain equal to the sum of the strains due to the various slip systems:

\be
\epsilon^0_{ij}({\bf r}) = \sum_{s:~slip~syst.} \epsilon^{00}_{ij}(s) \theta_s({\bf r})
\ee

\subsection{Coupling with a concentration field}

The main advantage of the present method is its ability to couple elastically in a straightforward way a dislocation field with any other field, and in particular with a concentration field.

We consider a coherent phase-separating alloy with a lattice parameter mismatch between two atomic species. The local concentration $c(\br)$ of the minority atomic species is taken as the field. The chemical free energy is represented by a coarse-grained Ginzburg-Landau free energy:

\be
F^{chem} = \int_V d^3r~(-\frac{\mu}{2}\phi^2+\frac{\gamma}{4}\phi^4+\frac{\lambda}{2} \| \nabla \phi \| ^2)
\ee
where $\phi(\br) = 2 c(\br) - 1$.

The total free energy of the system is then the sum of this chemical free energy and an elastic energy of the form of \Eq{EFinale}. This energy is related to the lattice parameter mismatch by means of a stress-free strain following Vegard's law : 
\be
\epsilon^0_{ij}({\bf r}) = \epsilon_0 \delta_{ij} c({\bf r})
\ee
where $\epsilon_0 = \frac{1}{a} \frac{\partial a}{\partial c}$.

Coupling with a dislocation field arises from the assumption that the total stress-free strain is the sum of the strains due to the dislocations and to the concentration field:

\be
\epsilon^0_{ij}({\bf r}) = \sum_{s:~slip~syst.} \epsilon^{00}_{ij}(s) \theta_s({\bf r}) + \epsilon_0 \delta_{ij} c({\bf r})
\ee

The elastic energy is given by \Eq{EFinale} where now the cross-matrices between dislocation and concentration fields characterize the elastic interactions between these fields.

\section{DYNAMICAL MODEL}

We now consider a method to apply dynamics to a dislocation field. In the case of a concentration or of a long-range order parameter field, dynamics are applied by means of Time-Dependent Ginzburg-Landau (TDGL) kinetic equations, which assume that the rate of change of the field is linearly proportional to its driving force, i.e. to the functional derivative of the total free energy of the system with respect to the field. This method is often referred to as a {\it Diffuse Interface approach} because density fields vary continuously along interfaces between phases, leading to interfaces with finite widths.

We present here dynamical equations for dislocation fields adapted from a Diffuse Interface approach. The dislocation fields take integer values everywhere in a slip plane, except along the dislocation lines where they vary continuously from one integer value to another, in regions analogous to dislocation cores.

\subsection{Dislocation Dynamics}

By analogy with a Diffuse Interface approach, dislocation fields are stabilized to integer values by associating to the fields a multi-well energy with minima every integer value. A similar approach is developed independently by Wang \etal \cite{KhachaActaMet}. This type of energy is reminiscent of the energy introduced in the Peierls-Nabarro (PN) model \cite{pn} to model the resistance of a crystal to slip. We note however that the two energies are different because (1) the present energy is defined in the space of dislocation fields and not real space and (2) the present energy need not (and in fact must not) reproduce the elastic response of the material near the bottom of the wells. In the PN model, the energy serves to glue two distinct elastic half-spaces together, whereas here, dislocation fields are superimposed to a continuous elastic medium. Fitting the multi-well energy on elastic constants would result, for example in presence of an external stress, in counting twice the elastic deformation of the material. The curvature at the bottom of the wells must therefore be large compared to the elastic constants.

Different expressions for this energy are possible. We have chosen a periodic function made of {\it Gaussian wells} which allows to choose separately the depth of the wells (related to the theoretical elastic limit of the material) and the curvature of the wells. The general expression is:
\be
E^{well} = A_w \int_V d^3r \sum_{s:~slip~syst.} 1-\exp(-\frac{H^2(\theta_s(\br))}{2 \sigma^2})
\ee
where $H(\theta)$ is the decimal part of $\theta$, with values between $-0.5$ and $0.5$.

As in classical Diffuse Interface methods, a gradient energy is added to stabilize the width of the interfaces, i.e. the width of the dislocation cores. This gradient energy must be non-zero only along the dislocation lines and must not penalize the jump in dislocation field across the loop surfaces. A suitable form for this energy is:

\be
E^{gradient} = \frac{A_g}{2} \int_V d^3r \sum_{s:~slip~syst.} \| {\bf n}_s \wedge {\bf \nabla} \theta_s(\br) \|^2
\ee

Dynamics are then applied to the dislocation fields by means of TDGL kinetic equations:

\be
\frac{\partial \theta_s}{\partial t} = -L \frac{\delta}{\delta \theta_s} \{ E^{el}+E^{well}+E^{gradient} \}
\ee
where $L$ is a kinetic coefficient fitted on dislocation mobilities.

The main advantage of this method is that it takes full advantage of the Diffuse Interface approach: dislocation positions do not have to be stored and short range reactions, such as annihilation between dislocations in a slip plane, are automatically accounted for and result simply in the merging of two interfaces.

The main disadvantage of this method is that it leads to largely spread dislocation cores: cores expand typically over 5 grid spacings, i.e. 50b with a spacing of 10b where b is the dislocation Burgers vector, which is large compared to typical core radii. Moreover, the maximum stress near a dislocation line being of the order $\frac{\mu}{ 2 \pi} \frac{b}{r_0}$ ($r_0$ is the core radius), the larger the core radius, the smaller the stress at short distances. This may lead to weak short-range interactions between dislocations and between dislocations and concentration fields. The core radii must therefore be kept as narrow as possible.

\subsection{Concentration field dynamics in presence of dislocations}

Classically, the dynamics of a concentration field is applied by means of a conserved TDGL kinetic equation with constant atomic mobilities. In the context of this approximation, dislocations produce a driving force on the concentration field independent of that field, which may lead to atomic fluxes even in regions of vanishing atomic concentration. The dependence of the atomic mobility on the concentration must therefore be taken into account. We thus use the mean-field expression for the atomic mobility and the conserved TDGL kinetic equation has the form:  
\be
\frac{\partial c}{\partial t} = {\bf \nabla}.~B c(1-c)~{\bf \nabla} \frac{\delta F^{tot}}{\delta c}
\ee
where $F^{tot}$ contains the chemical Ginzburg-Landau free energy and the elastic energy arising from both the concentration field and the dislocations. The elastic coupling between Eqs. 16 and 17 implies a dynamical coupling between the dislocation and the concentration field.

\section{APPLICATIONS}

We present in this section two applications of the method: simulation of a Frank-Read source in absence of a concentration field and shrinkage of a dislocation loop in presence of a concentration field.

In order to implement numerically the PFM, real space must be discretized. We have chosen a 3D square grid, the spacing of which sets the minimal thickness $d$ of the dislocation loops. Since this thickness strongly influences the maximum stress near the dislocation lines, $d$ must be of the order of the dislocation Burgers vector $b$. For that reason, we use a grid spacing $d = 10 b$. The other parameters entering the model are fitted on typical elastic constants, thermodynamic energies, atomic mobilities and dislocation mobilities.

\subsection{Frank-Read source}

An initial rectangular dislocation loop is placed in the central horizontal plane of a simulation cell containing $256 \times 256 \times 128$ grid points. Since the operation of a Frank-Read source consists in the bowing-out of a single dislocation segment, part of the initial loop must be neutralized which is obtained by placing a second dislocation loop of same Burgers vector as the initial loop in a plane perpendicular to the initial glide plane.

\begin{figure}
\centerline{\epsfxsize=1.5in \epsfbox{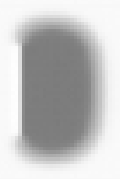} \hspace{0.1in} \epsfxsize=1.5in \epsfbox{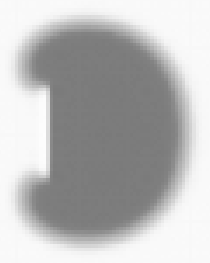} \hspace{0.1in} \epsfxsize=1.5in \epsfbox{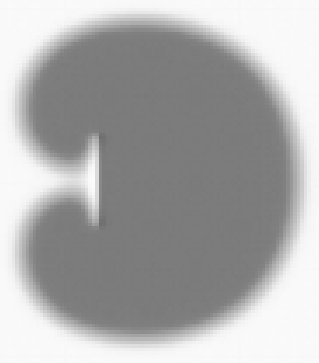} }
\vspace{0.1in}
\centerline{\epsfxsize=1.5in \epsfbox{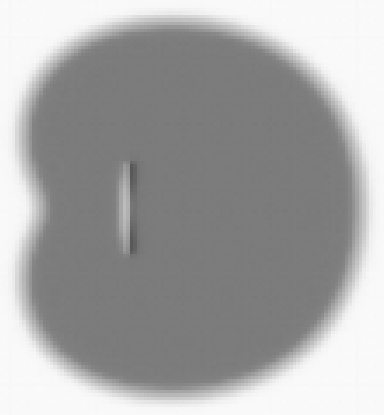} \hspace{0.1in} \epsfxsize=1.5in \epsfbox{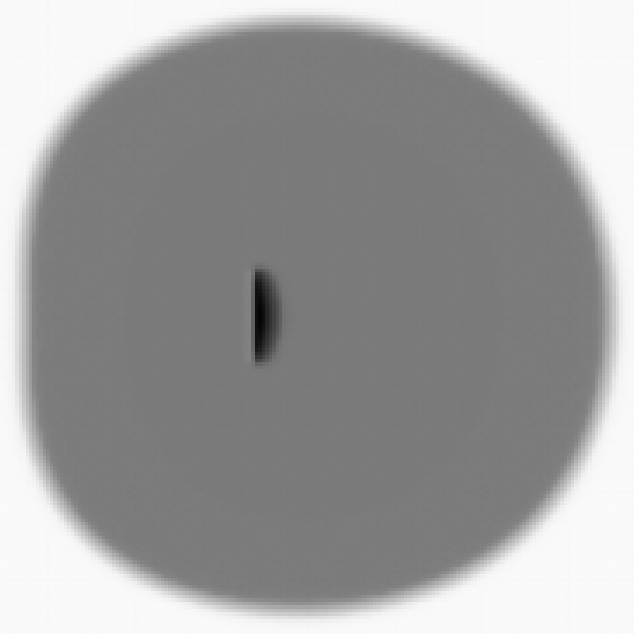} \hspace{0.1in} \epsfxsize=1.5in \epsfbox{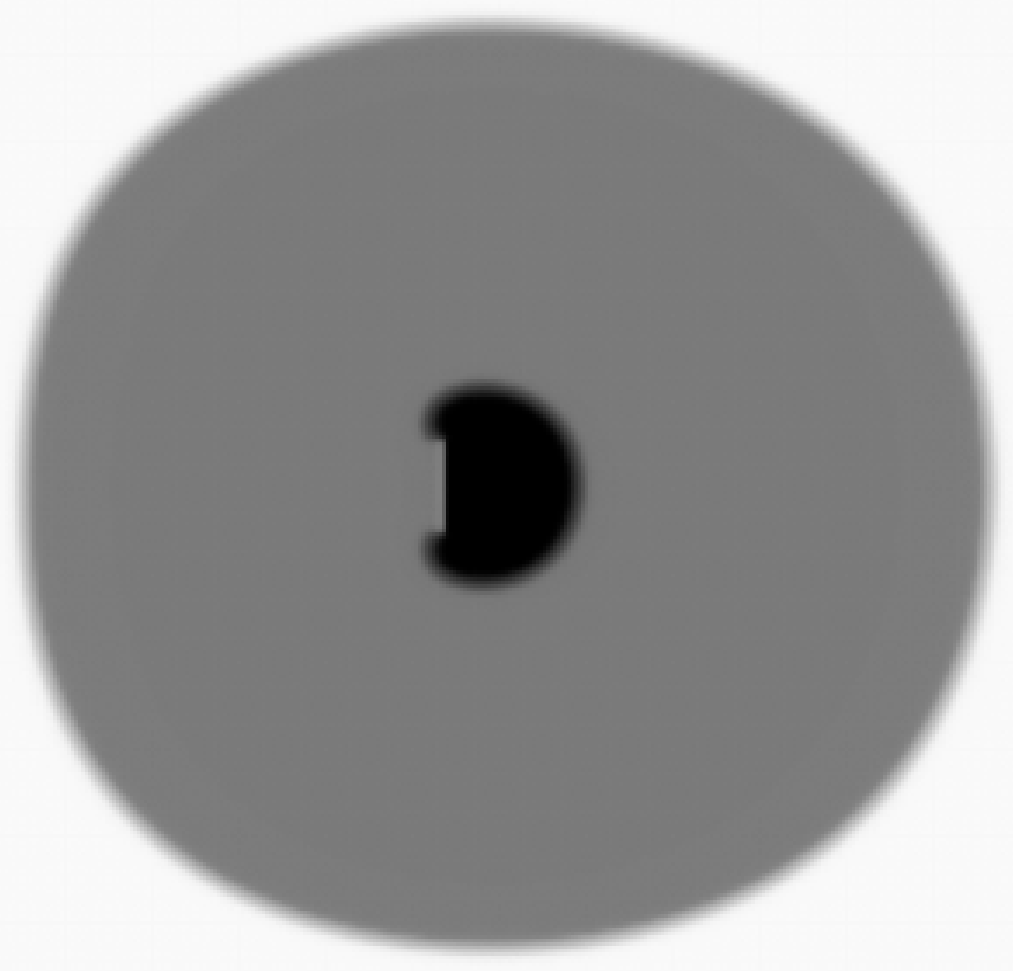} }
\medskip
\caption{Frank-Read source at different times. The color corresponds to the dislocation field, i.e. to the slip produced by the bowing dislocation. Darker colors correspond to larger slips.}
\label{FRsource}
\end{figure}

\Fig{FRsource} shows the motion of the pinned segment due to the application of a shear stress and clearly demonstrates the capability of the method to simulate an operating Frank-Read source.

\subsection{Shrinkage of a dislocation loop in presence of a concentration field}

\begin{figure}
\centerline{\epsfxsize=1.5in \epsfbox{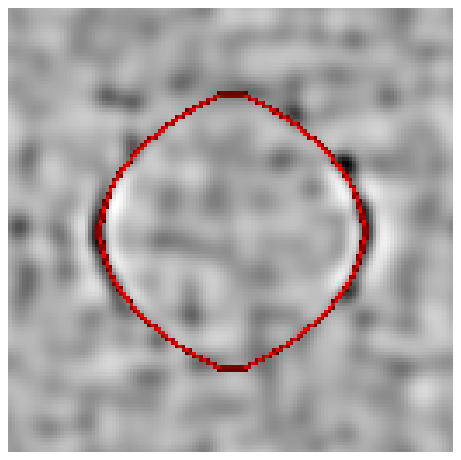} \hspace{0.1in} \epsfxsize=1.5in \epsfbox{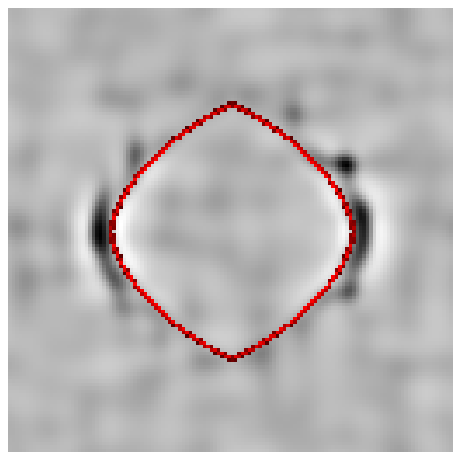} \hspace{0.1in} \epsfxsize=1.5in \epsfbox{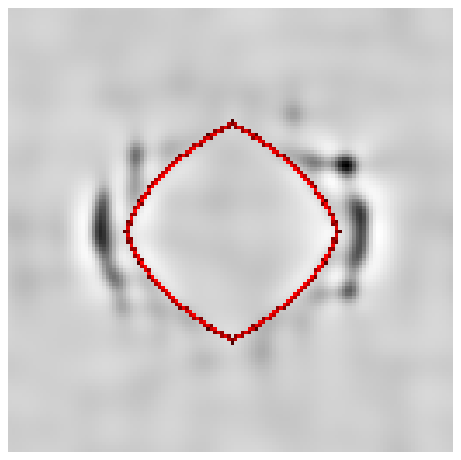} }
\vspace{0.1in}
\centerline{\epsfxsize=1.5in \epsfbox{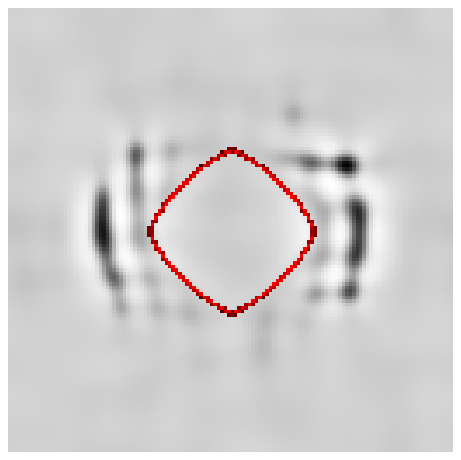} \hspace{0.1in} \epsfxsize=1.5in \epsfbox{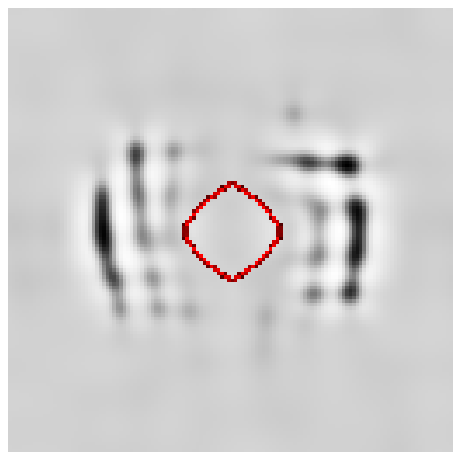}  \hspace{0.1in} \epsfxsize=1.5in \epsfbox{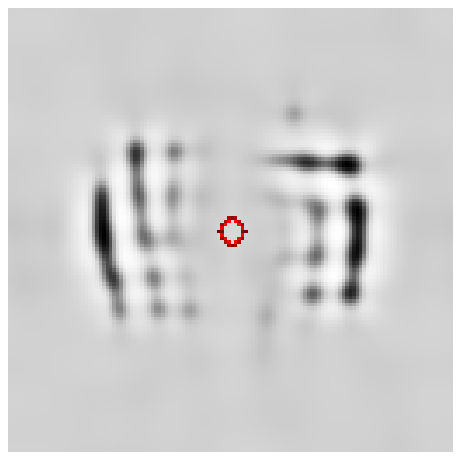} }
\medskip
\caption{Skrinkage of a loop in presence of an evolving concentration field.}
\label{shrinkage}
\end{figure}

We present in \Fig{shrinkage} the dynamics of a slip loop in presence of an evolving phase separating system initially disordered. The interaction between the dislocation strain field and the precipitates tends to lock the loop at the interfaces. In the present situation, the interaction between the loop segments is high enough to overcome this locking tendency and the loop area decreases. As a result, the gliding edge components of the loop leave in their wake two sequences of precipitates.

\section{Conclusion}

In summary, we have presented a general formalism for introducing dislocations in Phase Field methods. This approach can be used for any dislocation distribution in a 3-dimensional elastically anisotropic system. It has been illustrated here on the dynamics of a Frank Read source and of a phase separating system in presence of a moving loop, but can be used for more complex systems, where for example the structural inhomogeneities are represented by long-range order parameter fields. This gives us a new method for 3-dimensional mesoscopic simulations of the mechanical and plastic properties of complex microstructures. The main point of this method is that it automatically incorporates the interplay between precipitates and dislocation strain fields and, as a consequence, takes into account the interaction between the two dynamics.

\end{document}